\documentstyle[eqsecnum,aps,prl,twocolumn,mytitle]{revtex}

\input{psfig}

\begin{document}
\draft

\title{Hadroproduction of the $\chi_{1}$\ and $\chi_{2}$\ States of
Charmonium \\ in 800 GeV/c Proton-Silicon Interactions}

\author{\parindent=0.in
T.~Alexopoulos$^{18}$, L.~Antoniazzi$^{11}$, M.~Arenton$^{17}$,
H.C.~Ballagh$^1$, H.~Bingham$^{1}$, A.~Blankman$^{12}$, M.~Block$^{10}$,
A.~Boden$^2$, G.~Bonomi$^{11}$, Z.L.~Cao$^{17}$, T.Y.~Chen$^9$,
K.~Clark$^{15}$, D.~Cline$^2$, S.~Conetti$^{17}$, M.~Cooper$^{16}$, 
G.~Corti$^{17}$, B.~Cox$^{17,\dagger}$, P.~Creti$^7$, E.C.~Dukes$^{17}$,
C.~Durandet$^{18}$, V.~Elia$^7$, A.R.~Erwin$^{18}$, L.~Fortney$^4$,
V.~Golovatyuk$^7$, E.~Gorini$^7$, F.~Grancagnolo$^7$,
K.~Hagan$^{17}$, M.~Haire$^{13}$, P.~Hanlet$^{17}$, M.~He$^{14}$,
G.~Introzzi$^{11}$, M.~Jenkins$^{15}$, D.~Judd$^{13}$, W.~Kononenko$^{12}$,
W.~Kowald$^4$, K.~Lau$^6$, A.~Ledovskoy$^{17}$, G.~Liguori$^{11}$, J.~Lys$^1$,
P.O.~Mazur$^5$, A.P.~McManus$^{17}$, S.~Misawa$^1$, G.H.~Mo$^6$,
C.T.~Murphy$^5$, K.~Nelson$^{17}$, V.~Pogosyan$^{17}$,
S.~Ramachandran$^2$, J.~Rhoades$^2$, W.~Selove$^{12}$, R.P.~Smith$^5$,
L.~Spiegel$^5$,  J.G.~Sun$^{17}$, S.~Tokar$^3$, P.~Torre$^{11}$,
J.~Trischuk$^8$, L.~Turnbull$^{13}$, D.E.~Wagoner$^{13}$, C.R.~Wang$^{14}$,
C.~Wei$^{14}$, W.~Yang$^5$, N.~Yao$^9$, N.J.~Zhang$^{14}$, and
B.T.~Zou$^4$\\
(E771 Collaboration) \\
\vspace*{.1in}
\footnotesize
$^{1}$University of California at Berkeley, Berkeley, California, 94720 \\
$^{2}$University of California at Los Angeles, Los Angeles, California, 90024\\
$^{3}$Comenius University, Bratislava, Slovakia\\
$^{4}$Duke University, Durham, North Carolina, 27706\\
$^{5}$Fermi National Accelerator Laboratory, Batavia, Illinois, 60510\\
$^{6}$University of Houston, Houston, Texas, 77204\\
$^{7}$University and INFN of Lecce, I-73100 Lecce, Italy\\
$^{8}$McGill University, Montreal, PQ H3A 2T8, Canada\\
$^{9}$Nanjing University, Nanjing, People's Republic of China\\
$^{10}$Northwestern University, Evanston, Illinois, 60208\\
$^{11}$University and INFN of Pavia, I-27100 Pavia, Italy\\
$^{12}$University of Pennsylvania, Philadelphia, Pennsylvania, 19104\\
$^{13}$Prairie View A\&M, Prairie View, Texas, 77446\\
$^{14}$Shandong University, Jinan, Shandong, People's Republic of China\\
$^{15}$University of South Alabama, Mobile, Alabama, 36688\\
$^{16}$Vanier College, St. Laurent, PQ H4L 3X9, Canada\\
$^{17}$University of Virginia, Charlottesville, Virginia, 22901\\
$^{18}$University of Wisconsin, Madison, Wisconsin, 53706\\
\normalsize
}

\myabstract{
The cross sections for the hadroproduction of the 
$\chi_{1}$\ and $\chi_{2}$\ states of charmonium in proton-silicon 
collisions at $\sqrt{s}~=38.8~GeV$ have been measured in 
Fermilab fixed target Experiment 771.  The $\chi$ states were observed 
via their radiative decay to $J/\psi \gamma$, where the photon converted 
to $e^+e^-$ in the material of the spectrometer. The measured values for the
$\chi_{1}$\ and $\chi_{2}$ cross sections for $x_{F}>0$ are 
263~$\pm$~69(stat)~$\pm$~32(syst) and 498~$\pm$~143(stat)~$\pm$~67(syst)
nb per nucleon respectively. The resulting $\sigma(\chi_{1})/\sigma(\chi_{2})$ ratio of
0.53~$\pm$~0.20(stat)~$\pm$~0.07(syst), although somewhat larger than most theoretical 
expectations, can be accomodated by the latest theoretical estimates.
\vspace*{0.1in}
\noindent
}

\maketitle

\parindent=0.2in
\normalsize

Charmonium hadroproduction has provided interesting challenges to the 
understanding of QCD. Early attempts to describe the formation of a 
$c\overline{c}$ bound state, according to the \textit{color
evaporation}~\cite{cevap} or \textit{color singlet}~\cite{csing} models, did
not provide a satisfactory description of the available data. More recently,
the Non-Relativistic QCD Factorization Approach~\cite{bbl}, which incorporates
in a more rigorous fashion some of the features of the previous models, 
has provided a more successful description of the process. For any model, a 
rather crucial test has been the prediction of the relative rate of production for 
different charmonium states. In particular, when dealing with proton-induced 
processes, the absence of quark annihilation diagrams and the suppression 
in Leading Order of $\chi_{1}$ production gluonic diagrams implies rather
small values, typically less than 10\%, for the ratio of $\chi_1$ to $\chi_2$
production~\cite{thrat}. The measurement presented here represents a significant 
contribution to the available data, since it is the first observation 
of cleanly resolved $\chi_1$ and $\chi_2$ states in a proton-induced fixed 
target experiment.

The FNAL E771 experiment utilized a large-acceptance spectrometer~\cite{spect}
to measure several processes containing muons in the final state. Protons of 
800~GeV/c momentum were transported by the Fermilab Proton West beam line to
the High Intensity Laboratory, where they hit a 24 mm thick silicon target.
Operating at a beam intensity of $\approx 3.6~\times~10^{7}$ protons per spill-second, 
the experiment accumulated a total of 6.4~$\times~10^{11}$ p-Si interactions. 
The incoming proton beam trajectory and flux were measured by a 
six plane silicon detector station.  The 0.26 radiation length target 
was composed of twelve 2~mm silicon foils separated by 4~mm. The
target was followed by a microvertex detector consisting of fourteen 300~$\mu$m
thick silicon planes that, while not used in the analysis presented here, contributed an
additional 0.045 X$_0$ to the target region radiation length.

The spectrometer's tracking system consisted of seven multi-wire proportional 
chambers and three drift chambers upstream plus three drift chambers and 
six combination drift/pad/strip chambers downstream of a dipole analysis 
magnet which provided an 821~MeV/c $p_{t}$ kick in the horizontal plane.  
Downstream of the wire chamber system, an electromagnetic
calorimeter consisting of an active converter and 396 scintillating glass and 
lead glass blocks was used for electron/positron identification.  The final 
element of the spectrometer, a set of three planes of resistive plate counters 
(RPC's) segmented into 512 readout pads and sandwiched between 
layers of steel and concrete absorbers, provided muon identification.
The material in the absorber walls represented an energy 
loss of 10~GeV in the central region and 6~GeV in the outer region of the 
detector for the incident muons. 

A dimuon trigger\cite{dimutrig} selected events with a $J/\psi$\ in the 
final state via the decay $J/\psi \rightarrow \mu^{+}\mu^{-}$.
A trigger muon was defined as the triple coincidence of the 
\emph{OR} of 2~$\times$~2 pads in the first RPC plane and the \emph{OR} of 
6~$\times$~6 pads in the second and third RPC planes in projective 
arrangements.  A dimuon trigger was defined as two such triple coincidences. 
The trigger reduced the 1.9 MHz interaction rate by a factor of $\approx 10^4$, 
selecting approximately 1.3~$\times~10^{8}$ dimuon events to be written to tape.

The seed for muon track reconstruction was provided by the RPC 
triple coincidences.  The roads formed by the pads involved in 
the coincidences were projected into the rear chamber set, 
identifying a region in which to search for candidate muon tracks.  Muon 
tracks reconstructed in the rear chamber set were then matched with 
tracks found in the front chamber set by requiring a good
front-rear linking $\chi^{2}$.  Muon pairs were required to come from a
common vertex by applying a cut to their distance of closest approach.
About fifty thousand dimuon events survived the reconstruction process,
quality cuts, and vertex cuts.

\begin{figure}
 \begin{center}
  \parbox{8.0cm}{
  \psfig{figure= 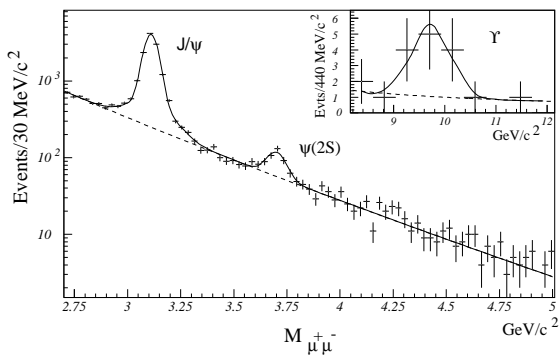,height=5.0cm}
  \caption{Opposite sign dimuon mass spectrum.  The solid line is a fit
           to the signal plus background and the dashed line is a 
           fit to the background only.  Inset shows the $\Upsilon$\ signal.}}
  \label{fig:dimuspec}
 \end{center}
\end{figure}

Figure~1 shows the resulting dimuon mass spectrum containing peaks
corresponding to the $J/\psi$, $\psi(2S)$, and $\Upsilon$\ (inset).  
Superimposed on the dimuon mass spectrum is a fit to the data obtained with 
the sum of two Gaussians for the $J/\psi$\ peak, a single Gaussian for the 
$\psi(2S)$, and the form $\frac{a}{m_{\mu\mu}^{3}}exp(-bm_{\mu\mu})$
for the continuum background.  The two Gaussians fit to the 
$J/\psi$\ peak is a good approximation (as confirmed by Monte Carlo) to a 
non-constant mass resolution, caused by the confusion associated with increases
in hit density near the beam region. The number of $J/\psi$'s
and $\psi(2S)$'s after background subtraction was 11,660~$\pm$~139 and 
218~$\pm$~24, respectively~\cite{psianal}.

Events in a window of $\pm$~100~MeV/c$^{2}$\ around the $J/\psi$\ mass were
refit varying the muon momenta within measurement errors, with the constraint
that the invariant mass of the pair be equal to $J/\psi$ mass. The resulting 
dimuon event sample was then inspected to search for $e^{+}e^{-}$ pairs that 
might be the result of conversions of photons from $\chi\rightarrow J/\psi~\gamma$
decays. Dimuon events which contained pairs of tracks matching the topology 
of a $\gamma \rightarrow e^{+}e^{-}$ conversion in the target region  -- 
collinear before the magnet in both bend and non-bend projections, collinear 
in the non-bend plane and coplanar in the bend plane after the magnet -- were 
then designated as $\chi$ decay candidates. 

All electron/positron pair candidates were required to satisfy additional conditions.
At least one of the two track candidates was required to be associated with 
an energy deposition in the calorimeter consistent with an electro-magnetic 
shower. In addition, the total transverse momentum of the $e^+e^-$ pair 
in the rear of the magnet was required to be zero (within the resolution of the
spectrometer) relative to the common $e^+e^-$ trajectory in front of the magnet.
To quantify how well a pair fitted the $\gamma \rightarrow e^{+}e^{-}$ hypothesis,
a $\chi^2$ was formed,

\begin{equation}
\chi^{2}=\frac{(a_{x1}-a_{x2})^2}{\sigma_{ax1}^{2}+\sigma_{ax2}^{2}} + 
         \frac{(a_{y1}-a_{y2})^2}{\sigma_{ay1}^{2}+\sigma_{ay2}^{2}} + 
         \frac{(b_{y1}-b_{y2})^2}{\sigma_{by1}^{2}+\sigma_{by2}^{2}}
\end{equation}

\noindent where $a_{x1}$ and $a_{x2}$ are the electron and positron track 
intercepts at the magnet in the bend plane, $a_{y1}$ and $a_{y2}$ the track
intercepts in the non-bend plane, $b_{y1}$ and $b_{y2}$ the track
slopes in the non-bend plane, and the $\sigma$'s are the measurement errors
on these quantities.  The electron/positron candidate with the smallest 
$\chi^{2}$\ in a given event was designated as a photon conversion
candidate.  Additional cuts requiring a good $\chi^{2}$, 
the transverse momentum of the parent photon to be between 250 and 700 MeV/c, 
and the invariant mass squared of the $e^{+}e^{-}$ pair 
to be less than 3000~(MeV/c$^2$)$^2$ were applied 
to maximize signal to background in the final sample of events containing both 
a $J/\psi$ and a photon conversion.

The $J/\psi~e^{+}e^{-}$ invariant mass shown in Fig.~2 was
calculated using the electron and positron momenta obtained from the 
tracking system.  
Clear $\chi_{1}$ and $\chi_{2}$ signals can be seen.  The background to 
the $\chi_{1}$ and $\chi_{2}$ was well-described by uncorrelated $e^+e^-$ 
and $J/\psi$ combinations: the solid line of Fig.~2 was obtained
by fitting two Gaussians plus a polynomial background.  The polynoimial
background was obtained by fitting
the mass distributions of $J/\psi$'s and $e^{+}e^{-}$'s extracted from different
events. The numbers of $\chi_{1}$ and $\chi_{2}$ obtained from the fit are
33~$\pm$~9 and 33~$\pm$~10, respectively. The fitted width is
5.2~$\pm$~2.0~MeV/c$^{2}$ for both the $\chi_{1}$ and $\chi_{2}$ peaks.  

\begin{figure}
 \begin{center}
  \parbox{8.0cm}{
  \psfig{figure= 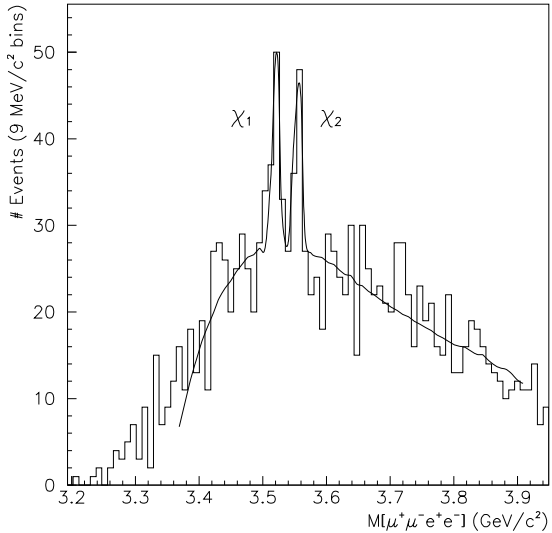,height=8.0cm}
  \caption{$J/\psi e^{+}e^{-}$ invariant mass.
           The solid line shows a fit to a polynomial background plus two
           Gaussians of equal width.  
           }}
  \label{fig:fit}
 \end{center}
\end{figure}

To determine the total cross section for $\chi_{1}$ and $\chi_{2}$ production,
the overall acceptance times efficiency for 
photon conversion and for electron/positron acceptance and 
recontruction efficiency had to 
be determined.   To accomplish this, a  Monte Carlo sample of 
$\chi \rightarrow J/\psi~\gamma$, $J/\psi \rightarrow \mu^{+}\mu^{-}$ decays 
was generated using Pythia~\cite{pythia}.  The photon and the muons were then
propagated through a GEANT~\cite{geant} simulation of the E771 detector, including
$\gamma$ conversion, scattering, bremsstrahlung and dE/dx. Hits from the Monte Carlo tracks 
obtained by this prescription were then inserted into actual dimuon trigger 
events to simulate realistically backgrounds and losses in pattern 
recognition due to confusion from noise hits and other tracks.  
Measured detector efficiencies were also applied to the inserted hits.
These hybrid Monte Carlo and data events were analyzed in a manner identical to the data in
order to determine acceptances and tracking efficiencies.  

Rather than attempting to simulate the electromagnetic calorimeter response 
to $e^{\pm}$ in detail in a Monte Carlo, the efficiency of matching an 
electron or positron candidate to a shower in the calorimeter was determined 
using a large sample of electron/positron pairs from photon conversions in 
minimum bias events. A sample of $e^{+}e^{-}$ pairs with kinematics similar to those
of the $\chi$ $e^+e^-$ pairs was collected using very 
tight cuts to ensure an $e^+e^-$ identity. This sample was
then subjected to the same constraints as those applied in the $\chi$ analysis.
The resulting overall acceptance times efficiency (inclusive of conversion probability)
for photons from $\chi$ decay was determined to be $(8.25 \pm 0.4) \times 10^{-3}$.

Using the $\gamma\rightarrow e^+e^-$ acceptance and efficiency, the measured 
branching ratios for $\chi_{1}$ and $\chi_{2}$ into $J/\psi~\gamma$~\cite{pdg}, 
the measured $J/\psi$ pN forward cross section~\cite{psianal} at 
$\sqrt{s}$=38.8 GeV and the number of observed 
$\chi_{1}$, $\chi_{2}$ and $J/\psi\rightarrow\mu^+\mu^-$, the absolute 
$\chi_{1}$ and $\chi_{2}$ cross sections for $x_{F}>0$ were
calculated to be $\sigma(\chi_{1})$ = 263 $\pm$ 69(stat) $\pm$ 32 (syst) 
nb/nucleon and $\sigma(\chi_{2})$ = 498 $\pm$ 143(stat) $\pm$ 67 (syst) 
nb/nucleon, respectively.  The main contributions to the systematic errors came 
from the error on the $J/\psi$ cross section (9\%), the uncertainty in the 
knowledge of the cut efficiencies (5\%) and the errors on the 
branching ratios for $\chi_{1}$ (6\%) and $\chi_{2}$ (8\%) .

Using the production cross sections for $\chi_{1}$ and 
$\chi_{2}$, the ratio of the $\chi_{1}$ to 
$\chi_{2}$ production cross sections was determined to be 
$\sigma(\chi_{1})/\sigma(\chi_{2})$ = 0.53 $\pm$ 0.20(stat) $\pm$ 0.07(syst).  
Combining this result with the two previous measurements of $\chi$ production  
by a proton beam~\cite{e673e705}, we have computed the world average (shown in Fig.~3) 
to be $\sigma(\chi_{1})/\sigma(\chi_{2})$ = 0.31 $\pm$ 0.14.
This figure is consistent with the latest NRQCD estimates of $\approx$0.3~\cite{fa},
where $\chi_{1}$ production was boosted by the inclusion of
higher order terms in the velocity expansion~\cite{gupta}.

\begin{figure}
 \begin{center}
  \parbox{8.0cm}{
  \psfig{figure= 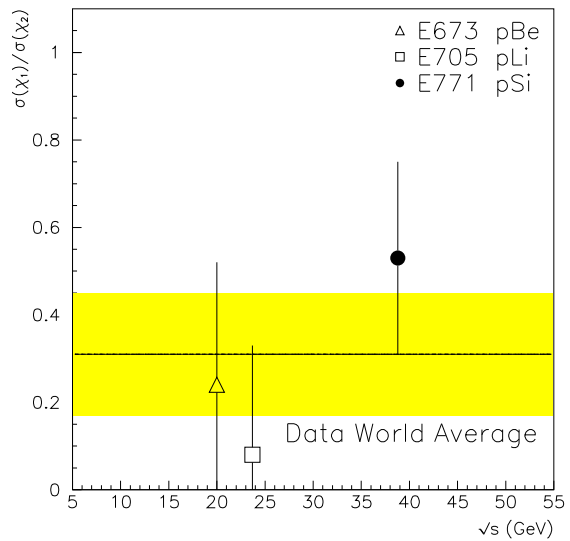,height=8.0cm}
  \caption{$\sigma(\chi_{1})/\sigma(\chi_{2})$ in proton-nucleon interactions.
           Results from this experiment (E771) are shown, together with the
           two existing results and the world average with 1$\sigma$ errors.}}
  \label{fig:chisigma_prl}
 \end{center}
\end{figure}

Finally, the energy dependence of the combined $\chi_{1}$ and $\chi_{2}$
production near threshold was compared to the corresponding quantity for
$J/\psi$ production. In Ref.~\cite{psidiff} data on $J/\psi$ production from
seventeen different pN experiments over a 
large range of center-of-mass energy, $\sqrt{s} \approx$ 8 to 52 GeV, 
were fit as a function of $\sqrt{s}$. The $J/\psi$ 
production data near threshold was well represented by the function 
$\sigma(\sqrt{s})_{J/\psi} = \sigma_{0}(1-M_{J/\psi}/\sqrt{s})^{\beta}$,
with $\sigma_{0}=1.0 \pm 0.1$
$\mu$b/nucleon and $\beta = 11.8 \pm 0.5$.  To check whether $\chi$ production 
has similar dynamics as $J/\psi$ production, the sum of the $\chi$ cross sections
have been fit 
to a similar parameterization with $\beta$ fixed to the $J/\psi$ 
value and $M_{\chi}$ replacing $M_{J/\psi}$. The result of the fit, shown in
 Fig.~4, shows the similarity of the $J/\psi$ threshold production 
parametrization to the threshold behavior of the combined 
$\chi$ state cross sections.  The fit yields $\sigma_{0}=2.3\pm 0.4\mu$b/nucleon 
for the asymptotic $\sigma(\chi)$ cross section.

\begin{figure}
 \begin{center}
  \parbox{8.0cm}{
  \psfig{figure= 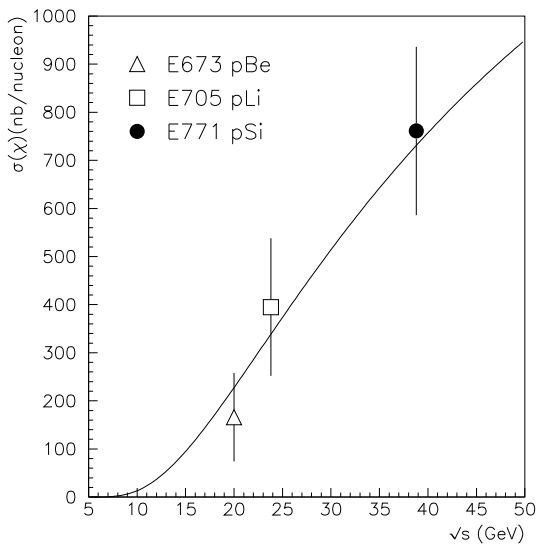,height=8.0cm}
  \caption{$\sigma(\chi)$ vs. $\sqrt{s}$ in 
           proton-nucleon interactions.  Superimposed on the data is a fit to
           a threshold production parameterization described in the text;  here
           $\sigma(\chi)$ is the sum of the $\chi_1$ and $\chi_2$ cross sections
           where the E673 data point has been obtained from the published $\chi_1$
           cross section and ratio of $\chi_2$ to $\chi_1$}}
  \label{fig:chi}
 \end{center}
\end{figure}

We wish to thank Fermilab, the U.S. Department of Energy, the National Science 
Foundation, the Istituto Nazionale di Fisica Nucleare of Italy, the 
Natural Science and Engineering Research Council of Canada, the Institute for 
Particle and Nuclear Physics of the Commonwealth of Virginia, and the Texas 
Advanced Research Program for their support.

\vspace{0.25cm}
\noindent $^{\dagger}$ To whom correspondence should be addressed.\\
\noindent Electronic address: cox@uvahep.phys.virginia.edu \\

\end{document}